\documentclass[a4paper]{jpconf}
\usepackage{graphicx}
\begin{document}
\title{Neutrinos, Rare Isotopes of Exotic Nuclei and Nuclear Astrophysics}

\author{A.B. Balantekin}

\address{Physics Department, University of Wisconsin, Madison, WI  53706 USA}

\ead{baha@physics.wisc.edu}

\begin{abstract}
The connection between neutrino physics, nucleosynthesis of elements in astrophysical sites, laboratory measurements with rare exotic nuclei and astronomical observations is discussed. The key role played by neutrinos is emphasized and the close connection between neutrino physics and nucleosynthesis is highlighted. 
\end{abstract}

Understanding where and how various nuclei are synthesized during the evolution of the Universe is one of the key questions of modern science. 
Element synthesis is thought to be a multi-site and multi-epoch process as depicted in Figure 1. 
Tackling the question of the origin of elements requires a multitude of tools: High-quality observations of stellar spectra, laboratory atomic physics data, modeling stellar photospheres as well as theoretical and experimental investigations of the relevant nuclear processes. In addition, typically copious amounts of neutrinos are present in most nucleosynthesis sites. This feature makes neutrino physics and neutrino-nucleus interactions salient components of many nucleosynthesis scenarios.  This very close link between astronomical observations, neutrino physics, and nucleosynthesis is depicted in Figure 2 and was described in detail in a recent special issue of Journal of Physics G where the reader is referred to for a more complete explanation of many details \cite{Volpe:2014yqa}.

In nuclear physics on the theoretical side there have been significant accomplishments during the last decade in calculations with microscopic ab initio methods, driven in part by advances in computational capabilities. The most salient of these results is the realization of the 
necessity of incorporating three-body forces, in addition to the conventional two-body forces,  to describe a wide range of nuclear phenomena. On the experimental side several facilities, already operating or under construction, with intense beams of rare isotopes now have the ability to explore nuclear physics far from stability, at the r-process nucleosynthesis 
path \cite{Balantekin:2014opa}. 

\begin{figure}[t]
\includegraphics[width=24pc]{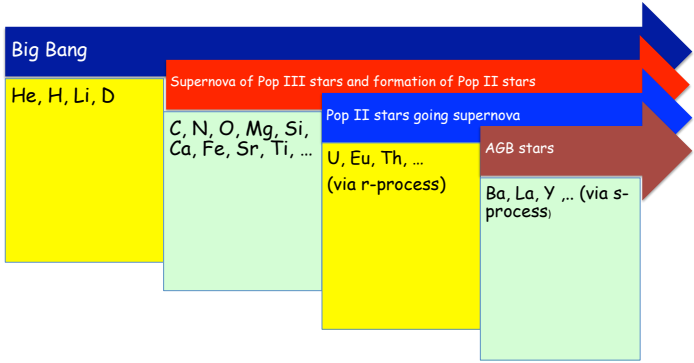}\hspace{2pc}%
\begin{minipage}[b]{10pc}\caption{\label{label}The sites of nucleosynthesis for various nuclei.}
\end{minipage}
\end{figure}

There is considerable debate about the astrophysical site of the r-process nucleosynthesis, the two most favorite sites being core-collapse supernovae and neutron-star mergers. Neutrinos not only play a crucial role in the dynamics of these sites, but they also control the value of the electron fraction, the parameter determining the yields in the r-process. (For a recent brief summary see Ref. \cite{Balantekin:2013gqa}). 
Understanding a core-collapse supernova requires answers to a variety of questions some of which need to be answered by nuclear physics, both theoretically and experimentally: Analysis of the formation and subsequent cooling of the neutron star involves high-density equation of state for nuclear matter. Studies of the oscillations of neutrinos as they travel from the neutrinosphere to where r-process or neutrino heating of the shock may take place need knowledge of neutrino properties. 
One should keep in mind that there are still many puzzles in neutrino physics. For example the near symmetry between 
second and third generations of neutrinos \cite{Balantekin:1999dx}, resulting from an almost maximal mixing angle $\theta_{23}$ and comparatively small mixing angle $\theta_{13}$ \cite{Balantekin:2013tqa}, seems to mimic astrophysical sites where muons and tau leptons 
do not exist in bulk and cannot be produced by neutrinos because of the energy considerations. We demonstrate this in Figure 3 where a mixing-matrix triangle ($|U_{\beta 2}|^2$ vs $|U_{\beta 3}|^2$) is plotted. (Note that since $|U_{\beta 1}|^2 + |U_{\beta 2}|^2 + |U_{\beta 3}|^2 =1$ for all $\beta = e, \mu, \tau$ only two of the mixing matrix entries are free parameters if one ignores possible sterile neutrinos that may mix with the active ones).   
Finally neutrino-nucleon and neutrino-nucleus interactions determine the electron fraction at the r-process site, element formation rate at the $\nu$-process site, and eventually detection rates at the terrestrial detectors. 

New and novel effects come into play when one considers neutrinos in a supernova. Energy released in a core-collapse supernova is 
about $10^{53}$ ergs which is $10^{59}$ MeV. 
99 \% of this energy is carried away by approximately $10^{58}$ neutrinos and antineutrinos. 
This necessitates including the effects of neutrino-neutrino interactions in the Hamiltonian describing neutrino propagation throughout the supernova:
\begin{equation}
\label{1}
H = \sum a^{\dagger} a + \sum (1- \cos \theta) a^{\dagger} a^{\dagger} aa , 
\end{equation}
where the first term describes neutrino oscillations as well as interaction of neutrinos with matter (including the MSW effect), and the second term describes neutrino-neutrino interactions. The second term makes the physics of a neutrino gas in a core-collapse supernova a very interesting and non-trivial many-body problem, driven by weak interactions. (The angle in the second term is the angle between momenta of the two interacting neutrinos), 
The resulting so-called collective neutrino oscillations gave rise to an active research area (for recent reviews see Refs. \cite{Duan:2009cd} and  \cite{Duan:2010bg}). 
Neutrino-neutrino interactions lead to interesting collective and emergent effects, such as conserved quantities and striking features in the neutrino energy spectra (spectral ÒswapsÓ or ÒsplitsÓ). Symmetries also seem to play an important role in delineating those phenomena 
 \cite{Duan:2008fd,Pehlivan:2011hp,Pehlivan:2014zua}. Of particular importance to the element production is the impact of collective neutrino oscillations on the electron fraction. In many, although not all, calculations of the collective neutrino oscillations the angle in Eq. (\ref{1}) is averaged over (called "single-angle" approximation). Indeed in the first calculation of the electron fraction resulting from the collective neutrino oscillations such an approximation was employed \cite{Balantekin:2004ug}. Subsequent work, however, highlighted the importance of "multi-angle" calculations \cite{Duan:2010af}. 

\begin{figure}[t]
\includegraphics[width=26pc]{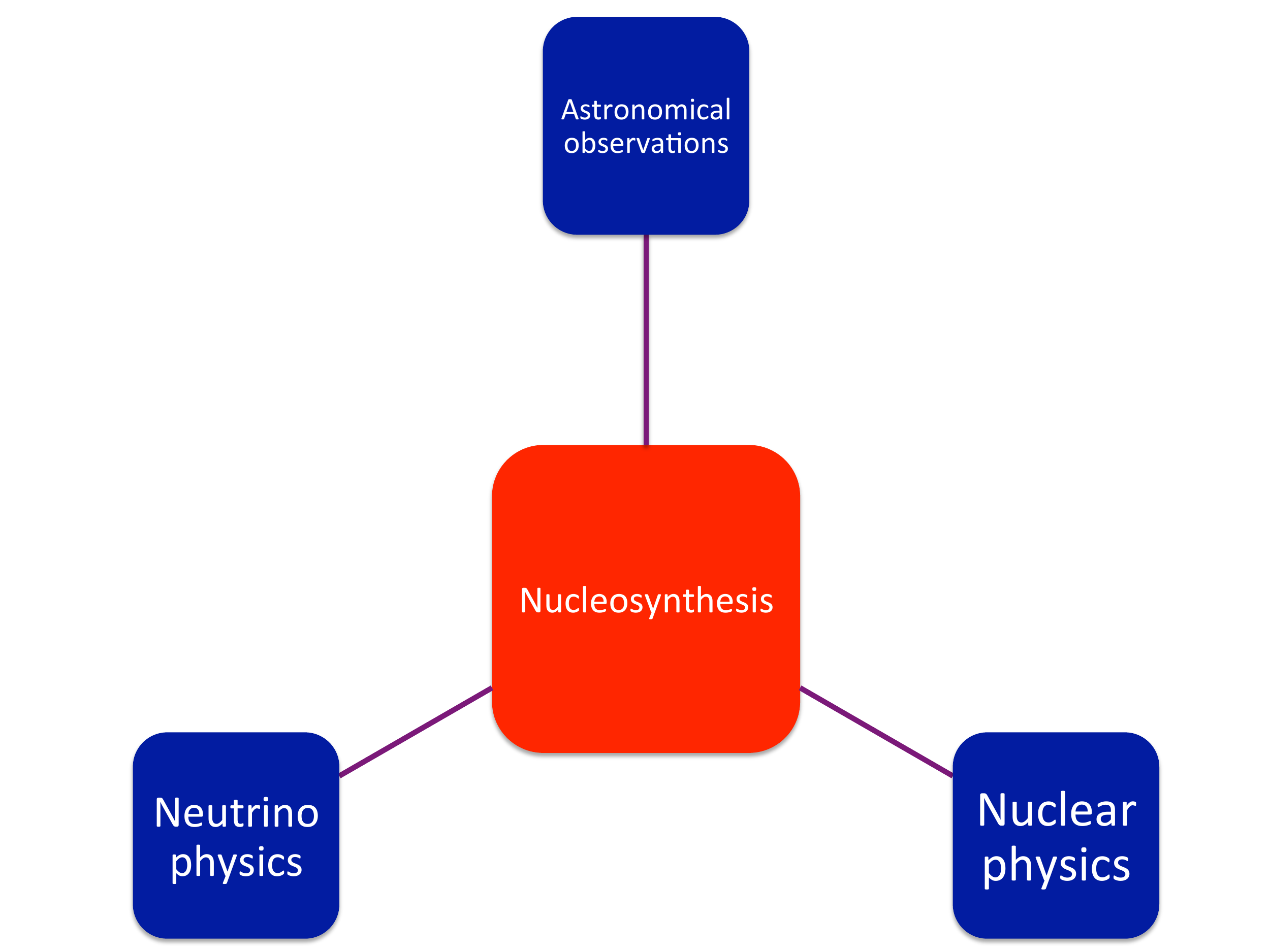}\hspace{2pc}%
\begin{minipage}[b]{10pc}\caption{\label{label}Three major physics input areas to describe nucleosynthesis.}
\end{minipage}
\end{figure}

It turns out that, because of practical limitations, calculations of the collective effects are carried out in an approximation where the neutrino in question 
travels in a mean field generated by the other neutrinos. The adequacy of this approximation is an open question, explored in Refs. \cite{Balantekin:2006tg} 
and \cite{Volpe:2013uxl}. However, at least in the single-angle case, the recognition that the invariants of the exact many-body Hamiltonian reduce to the appropriate constants of motion in the mean field limit 
\cite{Pehlivan:2011hp,Pehlivan:2014zua} suggests that the mean-field approximation appears to capture most of the relevant physics. 

It should be remarked that the astronomically observed fossil record of the r-process element abundances resulting from supernova explosions may also give us clues about the dynamics of the collapse event. For example if the progenitor star is sufficiently heavy, following the collapse a black hole may be formed at the center instead of a proto-neutron star. This would truncate emission of the neutrinos that cool the proto-neutron star, possibly altering the abundances of the elements formed during the r-process nucleosynthesis \cite{Sasaqui:2005rh}. Current data, however,  do not seem to support the possibility of the r-process truncation 
\cite{Aoki:2013eoa}. 

Part of the research program with exotic beams is to better understand the r-process nucleosynthesis. This requires understanding beta-decays of nuclei both at and far-from stability. One needs half-lifes at the r-process ladders (N = 50, 82, 126) where abundances peak, accurate values of initial and final state energies as well as the spin-isospin responses of the nuclei. Clearly it will not be possible for the exotic beam facilities to measure those quantities for all the nuclei in the r-process path; the task at hand is to identify particular measurements which will contribute the most to our comprehension of the r-process. 

Understanding the spin-isospin response of a broad range of nuclei to a variety of probes is crucial for astrophysics applications 
\cite{Balantekin:2014opa,Balantekin:2013gqa,Litvinova:2014jva,Cole:2012at}. For this purpose one needs to know not only the matrix elements of the Gamow-Teller operator $\mathbf{\sigma \tau}$ between the initial and final states, but, in a good number of cases, contributions from the forbidden transitions. Charge-exchange reaction experiments both with direct and inverse kinematics will help. 
Recently there have been significant developments in this area, particularly in the high-resolution study of the Gamow-Teller transitions with 
the ($^3$He, t)  reactions \cite{Fujita:2008zz,Fujita:2013wna}. On the theoretical side a persistent problem is the need to quench the axial-vector 
coupling strength $g_A$ in nuclei as compared to the free neutrons. Part of this quenching comes from the limited size of the model space and the effective interactions used. A recent development towards relaxing these limitations is a shell-model Hamiltonian, referred to as SFO. SFO Hamiltonian, enhancing monopole terms of the matrix elements in the $p_{1/2}$ and $p_{3/2}$ orbitals, includes tensor components consistent with the general sign rule for the tensor-monopole terms \cite{Suzuki:2003fw,Otsuka:2005zz,Suzuki:2007zza}. Calculations with this Hamiltonian reproduces the measured neutrino-$^{12}$C cross sections with a reduced quenching of $g_A$, as compared to the previous calculations \cite{Suzuki:2006qd}. Motivated by this success, SFO Hamiltonian is used to calculate neutrino-$^{13}$C cross sections
\cite{Suzuki:2014zea,Suzuki:2012aa}, which are helpful to know in analyzing data from carbon-based scintillators. 

\begin{figure}[t]
\includegraphics[width=36pc]{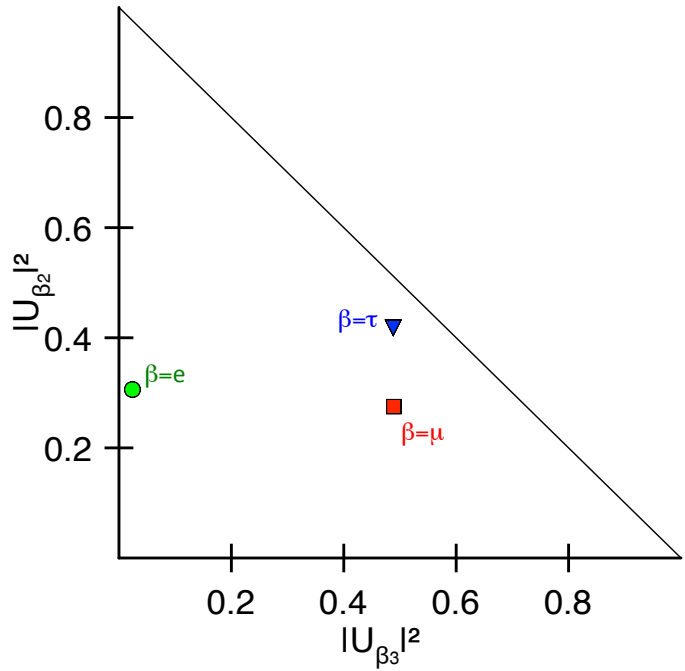}\hspace{2pc}%
\caption{\label{label}The mixing matrix triangle for the three flavors of neutrinos where $\beta$ is any of the  In this plot, ignoring any possible mixing between active and sterile neutrinos, the identity $|U_{\beta 1}|^2 + |U_{\beta 2}|^2 + |U_{\beta 3}|^2 =1$ is used. }
\end{figure}

Core-collapse supernovae are not the only astrophysical phenomena to explore which information about radioactive, unstable isotopes of nuclei is needed. A different kind of supernovae, called type IA, are set in motion by mass transfer onto the surface of a white dwarf from a giant companion star. This mass transfer is very rapid making white dwarf to approach Chandrasekhar limit and initiating a thermonuclear explosion. 
Sometimes the mass transfer is slow enough to cause explosive novae on the surface of a white dwarf star. Accreting neutron stars  from a binary companion can result in X-ray bursts similarly powered by nuclear reactions. To understand all those phenomena information to be gained from nuclear radioactive beam facilities is essential.

Although data radioactive-beam facilities are crucial to obtain a complete picture of element synthesis, one should not ignore the need for more precise data from stable-beam experiments. Arguably three most critical such reactions for nuclear astrophysics are $^3$He($\alpha, \gamma$)$^7$Be, $^{12}$C($\alpha,\gamma$)$^{16}$O, and $^{14}$N(p,$\gamma$)$^{15}$O \cite{Adelberger:2010qa}. The first reaction is the source of largest uncertainty in solar modeling and also Big Bang nucleosynthesis (albeit at different energy scales). The second reaction determines the ratio of $^{12}$C to $^{16}$O in the Universe, the determining parameter which makes the existence of living organisms for example on Earth possible. The third reaction is the key component of the CNO cycle. Although the Sun shines primarily with pp-cycle, CNO-cycle is the dominant energy generating mechanism for heavier stars, which comprises most of the past and present stars. 

Clearly many aspects  of the close connection between neutrino physics and nucleosynthesis still need to be explored using 
observational, theoretical, computational, and experimental tools available for a better understanding of the origin of elements. 

\ack
I would like to thank 
Akdeniz University nuclear physics group for its hospitality and 
my collaborators G. Fuller, W. Haxton, T. Kajino, Y. Pehlivan, and T. Suzuki for many enlightening discussions. 
This work was supported in part 
by the U.S. National Science Foundation Grant PHY-1205024 
and 
in part by the University of Wisconsin Research Committee with funds 
granted by the Wisconsin Alumni Research Foundation.

\end{document}